\documentclass[onecolumn]{aa}

\usepackage{graphicx}

\usepackage{txfonts}
\graphicspath{{./}{figures/}}
\usepackage[table]{xcolor}
\usepackage{amssymb,amsmath}
\usepackage{natbib}

\bibpunct{(}{)}{;}{a}{}{,} 
\usepackage{colortbl} 

\definecolor{gris}{rgb}{.8, .8, .8}

\usepackage{verbatim}

\begin{document}
\extrarowheight = -0.5ex
\renewcommand{\arraystretch}{2.25}

\title{A general relativistic estimation of the black hole mass-to-distance ratio at the core of TXS 2226-184}

\author{Artemisa Villalobos-Ram\'irez
          \inst{1}
          \and
          Oswaldo Gallardo-Rivera\inst{1}
          \and
          Alfredo Herrera-Aguilar\inst{1}
          \and
          Ulises Nucamendi\inst{2}
          }

   \institute{Instituto de F\'isica Luis Rivera Terrazas, Benem\'erita Universidad Aut\'onoma de Puebla\\
   CP 72570 Puebla, M\'exico.
              \\
         \and
             Instituto de F\'isica y Matem\'aticas, Universidad Michoacana de San Nicol\'as de Hidalgo, \\
             Edificio C–3, Ciudad Universitaria, CP 58040, Morelia, Michoac\'an, M\'exico.
             }
\abstract{In this work we make use of a general relativistic method to estimate the mass-to-distance ratio { ${M/D = 3.54^{+0.2}_{-0.2} } \times 10^4 M_\odot/Mpc$}  of the black hole hosted at the core of the active galactic nucleus of TXS 2226-184, along with its  {Right Ascension offset and the recession redshift (velocity)} of the galaxy. Our statistical fit is based on the frequency shift of photons emitted by water masers and their orbital positions when circularly revolving around the black hole center within the accretion disk of the active galactic nucleus. By taking into account a previously reported distance to the galaxy, we compare the result of the black hole mass fit to an estimate based on a mass-luminosity correlation. We find that the black hole mass at the core of TXS 2226-184 obtained with the aid of the statistical fit using the general relativistic method, ${M = 3.67 ^{+0.2}_{-0.2}} \times 10^6 {\mathrm M}_\odot$, is approximately $0.6$ times the black hole mass, $ M_{BH} = 6.24  ^{+ 3.6}_ {- 2.3 } \times 10^6 {\mathrm M}_\odot$, computed with the mass-luminosity correlation.}{}{}{}{}

\maketitle
\section{Introduction} \label{sec:intro}

Over 100 years ago Albert Einstein published his theory of General Relativity (GR) \citep{einstein1915}. Two months later, 
Karl Schwarzschild published a solution to Einstein's field equations \citep{Schwa} describing the gravitational field outside a spherically symmetric and static body. This solution is useful to describe the spacetime curvature generated by astrophysical objects such as stars and was later understood to describe a black hole (BH). A BH is a region of spacetime where the gravitational field is so strong that not even light can escape beyond the event horizon. A Schwarzchild BH is fully characterized by its mass. In 1963, Kerr constructed a solution that describes the gravitational field of a rotating BH \citep{Kerr} which is completely characterized by its mass and angular momentum. 

In recent years there has been important observational evidence of the existence of BHs. For instance, the observation of the stars orbiting with a very short period around Sgr A* in the center of our galaxy indicates the presence of a supermassive BH \citep{Ghez, Ghez2, genzel, genzel2, Ghez3, genzel3}, the detection of gravitational waves produced by BHs merger by LIGO-Virgo collaborations \citep{GW, GW1, GW3}, and the imaging of the M87 BH shadow by the EHT collaboration \citep{M87}.

Since BHs do not emit electromagnetic radiation, one way to study these enigmatic entities consists in observing their influence on stars, accretion disks, gas particles, etc., that orbit them. For certain astrophysical systems, the positions of the orbiting bodies and the redshift/blueshift of the photons they emit are available and can be measured. For this reason, several models that relate these observational quantities of the orbiting objects to the mass and the mass-to-distance ratio of the BH have been developed.

Water megamasers, emitting at 22 GHz, are astrophysical objects that have been found within Active Galactic Nuclei (AGNs), where they orbit central BHs \citep{claussen1984water}. The prefix ``mega'' refers to the intense luminosity emitted by the water maser in the AGN $(L > 10 L_{\odot})$ compared to luminosity of galactic masers $(L < L_{\odot})$ which are associated with star-forming regions \citep{genzel1977h2o}. 

Very Long Baseline Interferometry (VLBI) is an accurate technique for observing the positions and displacements of these maser features, as it gives us appropriate sub-milliarcsecond resolution for objects at (sub)parsec distances from the center of active galaxies. Telescopes like the Green Bank Telescope (GBT), the NRAO Very Long Baseline Array (VLBA)\footnote{The VLBA is operated by Associated Universities, Inc., under a cooperative agreement with the National Science Foundation which is a facility of the National Radio Astronomy Observatory (NRAO).}, the European VLBI Network (EVN)\footnote{The European VLBI Network (EVN) is a network of radio telescopes located primarily in Europe and Asia, with additional antennas in South Africa and Puerto Rico. Support for proposal preparation, scheduling, and correlation of EVN projects is provided by the Joint Institute for VLBI ERIC (JIVE); ERIC stands for the European Research Infrastructure Consortium.} 
 and the Effelsberg 100 m one \citep{reid2009megamaser, MCP2, MCP3}, provide us with observational data of positions and redshifts associated with the megamasers that allow us to estimate the mass of their central BH within an appropriate model.

In \citep{Hern}, the authors used a dynamic Keplerian model with relativistic corrections for masers features to fit the mass of the BH hosted at the core of the NGC 4258 galaxy. This modeling related the radial velocity of water megamasers to the observed redshift of the emitted photons by making use of the optical definition of the redshift. The Newtonian approach works accurately when the object is far enough from the gravitational source, but when the orbiting objects get close enough to the BH event horizon, the general relativistic effects become stronger and relevant. Thus when this happens, the redshift of the photons begins to have important general and special relativistic contributions.

A model for test particles orbiting a Kerr BH was presented at \citep{2015}. In this formalism, the influence of the BH on the curvature of spacetime was taken into account, and therefore the so-called gravitational redshift was included in the total redshift of photons emitted by particles orbiting the BH.

In \citep{Herrera_ngc}, the authors used a simplified version of the latter model and considered a Schwarzschild BH with water megamasers orbiting the AGN of NGC 4258. The authors estimated the mass-to-distance ratio of the BH at the galactic core of this galaxy as well as its peculiar redshift using a general relativistic approach. The peculiar redshift is related to the peculiar velocity of the galaxy with respect to the distant observer using the same optical definition. 

Another way to estimate the BH mass without using a model based on gravity resides on a correlation between the mass of the BH and the bulge luminosity $M_{BH}-L_{bulge}$ of the host galaxy \citep{M-L}, it is important to note that this correlation uses the K-band in the near-infrared instead of the visible spectrum. This correlation was initially based on the study of disk galaxies, but it works also  
for elliptical galaxies because they are morphologically equivalent to the bulge of disk galaxies.

One of the brightest known H$_2$O maser sources was discovered in the AGN of the TXS 2226-184 galaxy using the Effelsberg telescope \citep{FirstTXS}. The isotropic luminosity associated with this water maser is so large, $ L = 6100 \pm 900 L_\odot$ \citep{FirstTXS}, that it is called a ``gigamaser''. The distance to TXS 2226-184 is $D = 103.8571 \pm 0.2606$ Mpc\footnote{In \citep{kuo2018} the reported distance to TXS 2226-184 is $D=107.1$ Mpc, but since this distance has no associated uncertainties we can not use this result to estimate the value of the black hole mass with properly propagated errors.} as a result of applying Hubble's law to the reported systemic velocity $V_{lsr,radio}=7270\pm 18.24$ km s$^{-1}$ in \citep{Taylor_2002,Gigamaser} and assuming $H_0= 70$ km s$^{-1}$ Mpc$^{-1}$ \citep{kuo2018} (these authors did not report uncertainties taken into account for the Hubble constant value). TXS 2226-184 has been optically classified as an elliptical galaxy \citep{FirstTXS}. 
 
 In \citep{ball} the authors observed with the VLBA seven H$_2$O maser emission clusters in TXS 2226-184. The clusters were linearly distributed from northeast to southwest with position angle $= +25^o$ . The observational maser data have five redshifted maser features and two blueshifted maser features, where only one blueshifted maser was along the linear distribution (the most southwest maser feature), and the other one was about 7 mas southeast of the linear distribution. Given these features, the authors associated the distribution of the maser clusters with a parsec-scale, rotating disk, where the farthest blueshifted maser was situated completely outside the disk. However, the data reported in \citep{ball} did not provide absolute positions of the maser features.

In \citep{Gigamaser} the authors made new observations of the H$ _2$O gigamaser in TXS 2226-184 with the VLBA (one epoch) and the European VLBI Network (EVN; two epochs). The authors detected six maser features in epoch 2017.45 (VLBA), one in epoch 2017.83 (EVN), and two in epoch 2018.44 (EVN). In the data corresponding to epoch 2017.45 (VLBA), only one blueshifted maser feature was detected, while the other 5 maser features were redshifted with respect to the systemic velocity of TXS 2226-184. In addition, the authors provided absolute positions of the maser features. 
 
Most of the masers located in the accretion disks of supermassive BHs are megamasers. Estimates of BH masses using megamaser dynamics are of the order of ${10^6-10^7 M_{\odot}}$. So far no estimate of the mass of central BHs in galaxies hosting a gigamaser has been made, leaving the question of whether the high luminosity of the maser is related to a central BH mass of magnitude greater than ${10^6-10^7 M_{\odot}}$.

New observations of the gigamaser made with the VLBA have been published in \citep{Gigamaser}, and in principle, give us the opportunity to investigate whether there is a connection between its intense luminosity and the mass of the central BH.

\section{General relativistic model}
We assume a static and spherically symmetric spacetime, so we use the Schwarzschild metric (in natural units):
 {\small \begin{equation}
    ds^2=\dfrac{dr^2}{f}+r^2(d\theta^2+\sin{\theta}^2 d\varphi^2)-fdt^2,\quad \quad   f=1-\frac{2m}{r},
\end{equation}} where $m = GM/c^2$, and $M$ is the BH mass. 

The general relativistic model which we use was developed in \citep{Herrera_ngc}. We consider that massive test particles (photon sources such as stars, masers, and other bodies) follow a geodesic path. The geodesic motion of massive test particles is described by the 4-velocity $U^\mu=(U^t,U^r,U^\theta,U^\varphi)$ normalized to unity $U^\mu U_ \mu=-1$.

From this relation we obtain an equation similar to the conservation law of energy for a non-relativistic particle with energy $E^2/2$. For the special case of equatorial and circular orbits, the expression for the 4-velocity components simplifies since $U^r= 0 =U^\theta$. Therefore

 {\small \begin{equation}
    U^t =\sqrt{\dfrac{r}{r-3m}}, \quad \quad U^\varphi= \pm \frac{1}{r}\sqrt{\dfrac{m}{r-3m}},
\end{equation}}here the $\pm$ signs correspond to the angular velocity direction of the orbiting object.

The photons emitted by the test particles have a 4-momentum $k^\mu=(k^t,k^r,k^\theta,k^\varphi)$ and move along null and equatorial geodesics ($k^\mu k_\mu=0$), so that the emitted and detected frequencies can be written in terms of the parameters of the metric.

The Schwarzschild redshift and blueshift that emitted photons experience on their way from the source bodies towards a static observer, which stands far away from the BH ($U^{\mu}|_{d} \! = \! (1,0,0,0)|_{d};   r_{d} \! \rightarrow \infty $), read
 {\small \begin{align} 
       {1 + z_{Schw_{1,2}} \equiv 1 + z_{g}+ z_{kin_{\pm}}   \equiv \frac{\omega_e}{\omega_d} = \frac{(k_{\mu} U^{\mu})|_{e}}{(k_{\mu} U^{\mu})|_{d}} = \frac{(U^{t}- b{_{\mp}} U^{\varphi})|_{e}}{(U^{t}- b{_{\mp}} U^{\varphi})|_{d}} \approx (U^{t}- b{_{\mp}} U^{\varphi})|_{e}} ,
\end{align} }
where the subscripts $(1/2)$ correspond to $(+/-)$, the subscript $(e)$ refers to the emitter, $(d)$ to the detector and $\omega$ is the photon frequency; here $b_{\mp}$ is the light bending (deflection) parameter to the rigth/left of the line of sight, $z_{g}$ and $z_{kin_{\pm}}$ represent the gravitational and kinematic redshifts, respectively, and are expressed as:

 {\small \begin{equation}
   {b_{\mp}= \mp \sqrt{- \frac{g_{\varphi \varphi}}{g_{tt}}} = \mp \sqrt{\dfrac{r_{e}^{3}}{r_e-2m}} }, 
\end{equation}}

 {\small \begin{equation}
    z_{g}= \sqrt{\dfrac{r_e}{r_e-3m}}-1 , \qquad z_{kin_{\pm}}=\pm \sqrt{\dfrac{m r_e}{(r_e-2m)(r_e-3m)}},
    \label{zg-zkin}
\end{equation}}where the $\pm$ signs in Eq. (\ref{zg-zkin}) correspond to an approaching/receding object with respect to a far away observer, yielding the kinematic redshift $z_{kin_{+}}$ and blueshift  $z_{kin_{-}}$. 

{Hereinafter we use the approximation $\Theta \approx r_e/D$ for the angular distance between a given maser and the BH position, where $r_d = D$ is the distance from the detector to the BH. Then the gravitational and kinematic redshift become:}
 {\small \begin{equation}
    {z_{g}= \sqrt{\dfrac{1}{1-3\frac{m}{D\Theta}}}-1 }, \qquad 
    {z_{kin_{\pm}}=\pm \sqrt{\dfrac{\frac{m}{D\Theta} }{(1-2\frac{m}{D\Theta})(1-3\frac{m}{D\Theta})}}.}
\end{equation}}

\begin{figure}[ht!]
    \centering
    \includegraphics[scale=1.6]{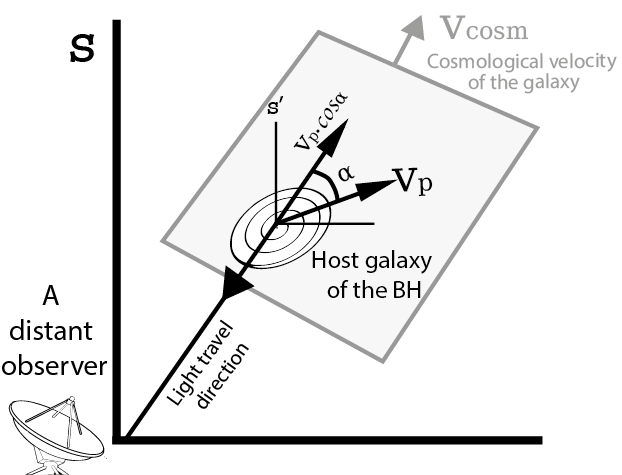}
    \caption{The motion of the galaxy hosting the black hole with respect to a distant observer. The galaxy has a peculiar velocity {${v_p}$ and a cosmological one ${v_{cosm}}$. The composition of the corresponding redshifts yields the observed recession redshift related to the velocity ${v_{rec}}$.}}
    \label{fig:doppler}
\end{figure}

For a full description of realistic systems, we must take into {account the recession redshift ${z_{rec}}$ given by the composition of the peculiar redshift, ${z_{p}}$, related to the peculiar velocity of the galaxy with respect to the observer, and the cosmological redshift, ${z}_{cosm}$, associated with the expansion of the universe  when the galaxy is within the Hubble flow; both of these redshifts have different nature.} Thus the total redshift is given by the following composition \citep{Composicion}\footnote{Strictly speaking one should consider a Schwarzschild-Friedmann-Robertson-Walker metric in order into taking into account the expansion of the universe in the BH geometry that generates the cosmological redshift. However we do not have the expression for the redshift for such a metric at hand, and as a first approximation we make use of the Schwarzschild background.\\

The peculiar redshift is defined through the special relativistic boost \citep{RindlerSR1989} 
 {\small \begin{equation*}
    1+z_{boost}=\gamma(1+ {\beta} \cos{\alpha}), \quad \gamma \equiv (1- {\beta^2})^{-1/2}, \quad {\beta\equiv \frac{v_p}{c}},
\end{equation*}}where ${v_p \equiv c~ z_p  }$, and ${v_p} \cos{\alpha}$ is the radial component of the peculiar velocity of the galaxy with respect to 
the observer (see Fig. \ref{fig:doppler}). Thus, in principle the $\alpha$-angle encodes the transversal motion of the galaxy with respect to the line of sight.}:
{\small \begin{equation}\label{redshi_tot}
    1+z_{tot_{1,2}}=(1+z_{Schw_{1,2}}){(1+z_{rec}); ~ (1+z_{rec})=(1+z_{boost})(1+z_{cosm}).}
\end{equation}}

\section{Mass-Luminosity correlation}
    Increased availability in BH mass demography has led to observe that there appears to be a close correlation between the BH mass and the bulge properties of the host galaxy. 
    The most studied correlations of bulge properties with black hole masses are given by: the mass-velocity dispersion correlation $M_{BH}-\sigma_{*}$ \citep{gebhardt2000relationship, ferrarese2000fundamental} and the mass-luminosity correlation $M_{BH}-L_{bulge}$ \citep{dressler1989active, kormendy1995inward, marconi2003relation}. In \citep{kormendy1995inward} the authors found that the BH masses show a correlation with the absolute blue luminosity of the host bulge for eight galaxies.

    In \citep{marconi2003relation} luminosity was taken in the $K-$band centered on $2.2 ~ \mu m$ (in the near-infrared 136 THz range) instead of the blue luminosity because the scatter in the first band is smaller than in the second one.  
 
    For an overview of the $M_{BH}-L_{bulge}$ correlation, followed by a larger sample of galaxies of different type and distinct BH masses hosted at their center, see Section 6 in \citep{M-L}. In the case of elliptical galaxies we must consider their total luminosity, because these galaxies are morphologically equivalent to the bulge component of disk galaxies. 
    
    The next equation shows the correlations of $M_{BH}$ with $L_{K,~bulge}$ \citep{M-L} 
  
 {\small \begin{equation} \label{lummm}
        \dfrac{M_{BH}}{ {10^{9}\mathrm M}_\odot}=\left( 0.542_{-0.061}^{+0.069}\right)\left(\dfrac{L_{K,~bulge}}{10^{11}L_{K\odot}}\right)^{1.21\pm 0.09},
    \end{equation}}where $L_{K,~bulge}$ is the bulge luminosity, and $L_{K\odot}$ is the solar luminosity, both in the $K$-band. 
 
\section{Observation of the H$_2$O gigamaser in TXS 2226-184}

In this paper, we consider the VLBI observations of gigamaser features in the AGN of TXS 2226-184. We use the data reported by \citep{Taylor_2002, Gigamaser}. The latter authors measured the redshift of photons emitted at the points of maximum emission and their absolute positions with errors lesser than 1 milliarcsecond. According to the authors \citep{Taylor_2002, Gigamaser}, the TXS 2226-184 galaxy is located at a distance $D= 103.8571 \pm 0.2606$ Mpc with adopted center at $\alpha_{2000} =22^h:29^m:12^s.494600 \pm 0.000291, \delta_{2000} =-18^h:10^m:47^s.24200 \pm 0.000409$, with peculiar velocity V$_{lsr,radio} = 7270 \pm 18.24$ km s$^{-1}$. 
We use the observations of redshift and masers positions corresponding to VLBA data epoch 2017.45 on June 12, 2017. The VLBA provides an angular resolution of $0.2$ miliarcsec (mas) and spectral resolution of 1 km  s$^{-1}$ at 22 GHz. The observational data of the maser were sparse, with only six masers features reported, five redshifted, and only one blueshifted. Despite the minimal quantity of data, we still can perform a statistical fit to estimate the mass-to-distance ratio of the central BH of this galaxy.

\section{Statistical fit with our general relativistic model}
The observations indicate that set of water maser clouds is allocated on the accretion disk of a central BH hosted at the AGN of the galaxy TXS 2226-184.  These features lie on the equatorial plane since we see the disk edge-on, and we shall assume that their motion is circular around the BH. Therefore we can make use of equation \eqref{redshi_tot} to model their total redshifts and blueshifts, which are directly observed.

To fit the parameters, we use the least-squares estimation $\chi^{2}$ by a Bayesian statistical ﬁt based on the Markov-Chain Monte Carlo scheme applied to the maser data using the general relativistic formalism. We emphasize that we apply our ﬁts to directly measured general relativistic invariant quantities.

\begin{figure}[ht!]
    \centering
    \includegraphics[scale=.24]{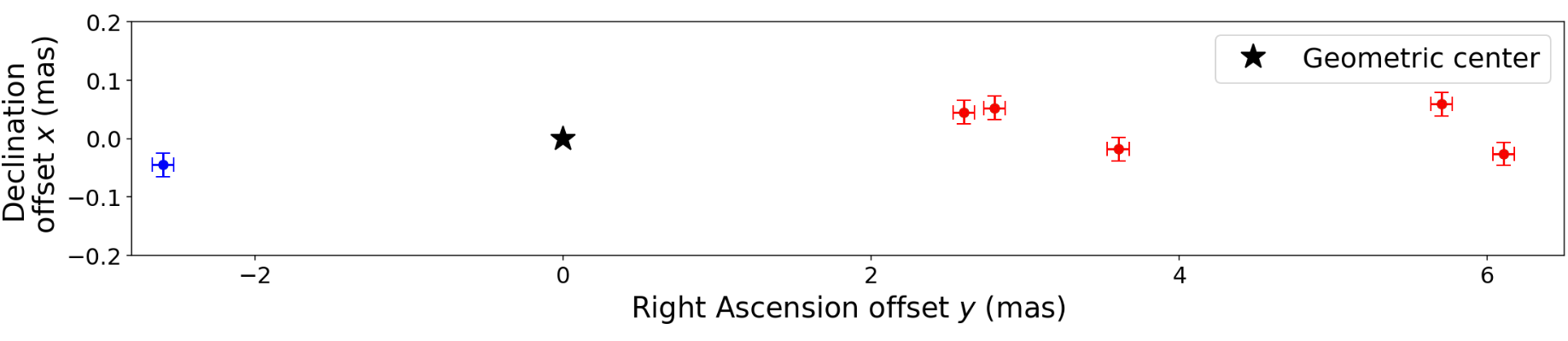}
    \caption{{\small View of the rotated H$_2$O maser system in TXS 2226-184. The star indicates the origin, the blue dot the blueshifted maser, and red dots the redshifted masers \citep{Gigamaser}.}}
    \label{fig:position}
    \end{figure}

{The parameters we fit are the mass-to-distance ratio M/D, the Right Ascension (RA) offset ${x_0}$ of the BH and the  recession redshift of the galaxy ${z_{rec}}$. The position data of the maser features reported in \citep{Gigamaser} are presented by taking the brightest maser as a reference.} Instead of taking a reference maser we propose a new reference origin at the geometric center of the maser system (see Fig. \ref{fig:position}), so that the fitted BH position will be estimated with reference to that point. {However, by varying $y_{0}$ within the observed height of the disk, we see that the estimate of $M/D$ changes at the third significant figure after the decimal point in comparison to the estimation with $y_0=0$ mas. This change is well behind the $M/D$ uncertainty and reveals the thin character of the disk, implying that $y_0$ does not influence the estimation of this quantity.} Indeed we shall assume that masers do not lie completely along the midline but are uniformly scattered about it with a scattering angle $\delta \varphi$  {and that the disk inclination $\theta_0$ is parameterized by the polar angle towards the equatorial plane.}

Now, we present the $\chi^2$ of the general relativistic model based on \citep{Hern, Herrera_ngc}:
{\small
\begin{equation}\label{chi}
\chi^2=\sum_{k=1}  \frac{\left[\frac{v_{k,obs}}{c}-(1+z_{g}+ \epsilon { \sin {\theta_0}} ~  z_{kin_{\pm}}){(1+z_{rec})}+1\right]^2 }{\sigma_{z_{tot_{1,2}}}^{2} + { {\kappa}}^{2} z_{kin_{\pm}}^{2} \sin^{2}{\theta_0}{(1+z_{rec})^2} },
\end{equation}}where the first term in the numerator refers to the observed redshift and the remaining terms are related to our model. In the denominator $\sigma_{z_{{tot}_{1,2}}}^2$ is the error associated with the total redshift. This quantity is $|\delta z_{{{tot}_{1,2}}}|^2$ and means the variation of the total redshift as shown in \citep{Herrera_ngc}
 {\small \begin{equation}
     \delta z_{tot_{1,2}}=(\delta z_{g} + \delta z_{kin_{\pm}})(1+z_{rec}).
 \end{equation}}

Following the latter work, we consider the redshift errors caused by the errors in the positions so that 
{\scriptsize$ \delta z_{g}=\left(1+z_{g}\right)^{3}\left[\dfrac{-3M}{2r^2}\right]\delta r$}, {\scriptsize $\quad \delta z_{kin_{\pm}} =\epsilon \sin{\theta_0}\left(z_{kin_{\pm}} \right)^{3}\left[\dfrac{6M^2-r^2}{2Mr^2}\right]\delta r$ } where {\scriptsize$\delta r = \sqrt{ \left( \frac{x_i-x_0}{r}\right)^2 \delta_{x}^2 + \left(\frac{y_i-y_0}{r}\right)^2 \delta_y^2}$}, and ($x_i$, $y_i)$ is the position of the i-th megamaser on the sky and $\delta x$, $\delta y$ are their respective errors.

 The quantities $\epsilon$, ${ {\kappa}}$ refer to the spread of the maser features by the azimuth angle \citep{Hern}

 {\small    \begin{equation}
        \epsilon \approx 1 - \frac{\delta \varphi^2}{2} + \frac{\delta \varphi^4}{24}, \quad \quad { {\kappa}}^2 \approx \frac{\delta \varphi^4}{4},
    \end{equation}}where the ﬁrst expansion corresponds to the cosine function of the azimuthal angle $\varphi$ and ${\kappa}$ denotes induced uncertainties of the maser scattering under the assumption that $\varphi \ll 1$ and $ \varphi \sim \delta \varphi$. 
    
    The observed data were rotated by an angle of $-87^o$ to fit the positions on the {horizontal axis}, and we considered an error propagation related to the rotation, according to the formula:
   {\small 
     {\small 
     \begin{align*}
        \sigma_{x_{rot}}^{2} = \left( \frac{ \partial x_{rot} }{\partial x} \sigma_x\right )^2 + \left(\frac{ \partial x_{rot}}{\partial y} \sigma_y \right)^2, \\
        \sigma_{y_{rot}}^{2} = \left( \frac{ \partial y_{rot} }{\partial x} \sigma_x\right )^2 + \left(\frac{ \partial y_{rot} }{\partial y} \sigma_y \right)^2,
    \end{align*} }}where $x_{rot},~ y_{rot}$ are the rotated positions, $\sigma_{x,y}$ are the non-rotated uncertainties, and $\sigma_{x_{rot}, y_{rot}}$ are the rotated ones.
    
\section{Results}

We have obtained two different estimates for the mass of the BH in TXS 2226-184 based on two different approaches being consistent between them. The estimate based on the motion of the masers and the frequency shift of the photons they emit is more precise. It is the first reliable estimate of the mass-to-distance ratio of the central BH of this galaxy.

\subsection{General relativistic model}

\begin{figure}[ht!]
        \centering
        \includegraphics[scale=.33]{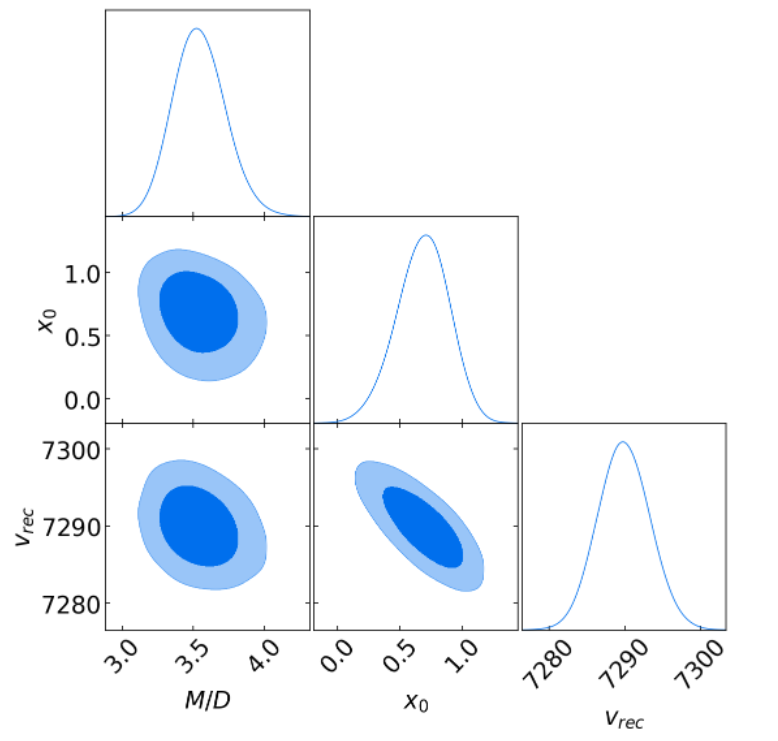}
        \caption{{\small The posterior distribution of the general relativistic Bayesian ﬁt. Here the BH mass-to-distance ratio $M/D$ is expressed in ${\times 10^4}$  \( M_\odot\)/Mpc, ${x_0}$  is expressed in mas, and{ $ {v_{rec}}$ in km s$^{-1}$. Contour levels correspond to 1$\sigma$ and 2$\sigma$ confidence regions. We have considered flat priors for $M/D$: [0, 200] ${\times 10^4}$  \( M_\odot\)/Mpc, $x_0: [-1.5 , 1.5]$ mas, and $z_{rec}: [0, 0.03]$ ($v_{rec}: [0 , 9000]$ km s$^{-1}$).
}}}
    \label{fig:posterior}
    \end{figure}

This method allows us to make a statistical fit with good accuracy of the mass-to-distance ratio (see Table \ref{t.result}) despite the fact that we have few data from the masers. We also find that our model provides a good fit to the data, yielding a value of the reduced ${\chi^2_{red}=1.512}$ with a maser scattering angle ${\delta \varphi =0.35}$ rad { and assuming a completely edge-on view of the accretion disk ($\theta_0 = \pi/2$ rad}). Table \ref{t.result} shows the values for the best fit and the uncertainties {with $1\sigma$ confidence} associated to each of the estimated parameters. In Fig. \ref{fig:posterior} we present the posterior distribution of the general relativistic Bayesian fit {with flat priors for the parameters of} the central BH at the core of TXS 2226-184.

    \begin{table}[ht!]
    \centering
    \caption{Posterior parameters for the BH located at the core of TXS 2226-184.}
    \label{t.result}
    \begin{tabular}{c c}
    \hline \hline
      TXS 2226-184  & Relativistic Estimation   \\ \hline \hline
      
      \rowcolor{gris}
      $M/D$  ($10^4$  \( M_\odot\)/Mpc)  & {\small ${3.54^{+0.2}_{-0.2}}$}  \\ 
     $z_{rec}$ $(10^{-2})$ & ${2.43_{-0.0011}^{+0.0011}}$  \\
     ${v_{rec}}$ ($km/s$) & {\small ${7289.8^{+3.4}_{-3.3}}$ }  \\ 
     \rowcolor{gris}
     $ x_0 $ (mas)  & ${0.69^{ +0.2}_{ -0.2}}$ \\
     $\chi^{2}_{red}$   &  {1.512} \\ \hline \hline
    \end{tabular}
    \end{table}
    
     In Table \ref{t.result} we also display the most probable value of the redshift ${z_{rec}}$ { (and its associated velocity ${v_{rec}}$) which is related with the recessional motion} of the galaxy. {The recession velocity} we obtain is consistent with the systemic velocity of the galaxy previously reported in \citep{Taylor_2002}. {Actually, what the statistical fit estimates is the recession redshift which is the composition of the cosmological and the peculiar redshifts, despite the fact that these quantities have a very different nature. Since both of these redshifts do not depend on the position of the masers, they are degenerate and cannot be estimated separately.
}

    If we use the distance to the BH based on \citep{Taylor_2002} {\small D= 103.85 $\pm$0.26 Mpc} we get a BH mass estimate:
 {\small \begin{equation}\label{masaRG}
       {M = 3.67 ^{+0.2}_{-0.2} \times 10^6 {\mathrm M}_\odot.}
    \end{equation}}
    
    We can also use this model to calculate the gravitational redshift of each maser. Below in Table \ref{t.zg} we display the gravitational redshift for the two closest masers to the black hole.\\
\begin{table}[ht!]
    \centering
    \caption{Gravitational redshift for the masers closest to the central BH.}
    \label{t.zg}
    \begin{tabular}{c c c}
    \hline \hline
         Maser & $z_g$ & velocity (km/s)\\ \hline \hline \rowcolor{gris}
         Red & {\small ${2.74}\times 10^{-7}$} & {\small ${8.21}\times 10^{-2}$} \\ 
          Blue & {\small${1.59}\times 10^{-7}$} & {\small ${4.77}\times 10^{-2}$}\\ \hline \hline
    \end{tabular}
\end{table}

    \subsection{BH Mass-Luminosity correlation}
    
    In \citep{FirstTXS}, the authors measured the luminosity of the TXS 2226-184 galaxy in the K-band, the adopted luminosity is $L_{k}= 2.5 \times 10^9 L_{\odot}$ with no reported uncertainties. In this framework, we accordingly substitute this reported luminosity into equation (\ref{lummm}), thus, the BH mass estimate reads
 {\small    \begin{equation}\label{masaML}
       M_{BH} = 6.24  ^{+ 3.6}_ {- 2.3 } \times 10^6 {\mathrm M}_\odot.
    \end{equation}}
    
    Using the mass-luminosity method, we get a mass of the same order of magnitude as the estimate using the method of general relativity, but we get a substantial uncertainty compared to that estimate, in fact an order in magnitude larger.
    
    Comparing these results, we see that a fit based on the redshift of the particles is more precise and reliable than using estimates based on the galaxy properties like luminosity. 
    
\section{Conclusions and discussion}

Our general relativistic approach provides estimates for the mass-to-distance ratio of the BH hosted at the AGN of TXS 2226-184 ({\small M/D=${3.54^{+0.2}_{-0.2}} \times 10^4  M_\odot/$Mpc}) as well as for its {RA offset, the recession redshift of} the host galaxy and its associated velocity. Furthermore, this model also allows us to quantify the gravitational redshift of each of the maser features; we calculate it for the two closest masers to the central BH of this galaxy. {The gravitational redshift obtained for each of the gigamasers is one order of magnitude smaller than the detector sensitivity, implying that this quantity cannot be currently detected in an unambiguous manner in this astrophysical system.}

Starting from our estimate of the mass-to-distance ratio of the BH located at the core of TXS 2226-184 and the distance to this galaxy based on a previous work \citep{Taylor_2002}, we obtain ${M = 3.67 ^{+0.2}_{-0.2}} \times 10^6 {\mathrm M}_\odot$. Therefore, TXS 2226-184 hosts a black hole with a mass of the same order expected for BHs hosted in galaxies associated with megamaser emission. This result allows us to conclude that the high luminosity of the gigamaser is not related to a more massive central black hole.

By comparing the results obtained for the mass of the BH hosted in TXS 2226-184, (see Eq. (\ref{masaRG}) and Eq. (\ref{masaML})), we find that the mass obtained from the $M_{BH}-L_{K, ~ bulge}$ correlation is approximately 1.6 times the mass obtained from the statistical fit using the general relativistic method. Finally, we note that the accuracy of the results differs by an order of magnitude, with the relativistic fit being the most accurate; however, these results are not mutually exclusive due to the uncertainties in the estimate based on the mass-luminosity correlation. Although less accurate, the $M_{BH}-L_{bulge}$ correlation is a good first approximation for systems for which there is no relevant data to make use of the general relativistic model. 

{Among the possible systematic errors of our modeling one could consider modifications of the edge-on view and a warped disk, in particular. By performing small variations in the inclination disk parameter $\theta_0$ (up to 5 degrees\footnote{According to \citep{MCP3} and \citep{Darling_2017}, if a thin disk is inclined by more than $\sim5$ degrees from an edge-on view, the masers will be no longer beamed toward us.}), the $M/D$ estimation gets changes of the order of 1\%, which is well-behind the uncertainty of this ratio. According to Eq. (\ref{chi}) the warping of a disk is correlated with the $M/D$ parameter. By considering a linear inclination warping along the radius of the disk, and performing variations in the inclination gradient (up to 0.04 rad/mas based on the masers distribution), the M/D estimation is altered around 5\% and the corresponding reduced $\chi^2_{red}=1.43.$}

\section*{Acknowledgments}
The authors are grateful to D. E. Villaraos-Serés for fruitful discussions and to FORDECYT-PRONACES-CONACYT for support under grant No. CF-MG-2558591; U.N. also was supported under grant CF- 140630. A.H.-A. and U.N. thank SNI and PROMEP-SEP and were supported by grants VIEP-BUAP No. 122 and CIC-UMSNH, respectively. O.G.-R. and A.V.-R. acknowledge ﬁnancial assistance from CONACYT through PhD grants No. 885032 and No. 1007718, respectively.

\bibliography{TXS_2226-184_black_hole_mass}{}

\begin{thebibliography}{35}
\expandafter\ifx\csname natexlab\endcsname\relax\def\natexlab#1{#1}\fi

\bibitem[{Abbott {et~al.}(2016)Abbott, Abbott, Abbott, Abernathy, Acernese,
  Ackley, Adams, Adams, Addesso, Adhikari, {et~al.}}]{GW}
Abbott, B.~P., Abbott, R., Abbott, T., {et~al.} 2016, \prl, 116, 061102

\bibitem[{Abbott {et~al.}(2020{\natexlab{a}})Abbott, Abbott, Abraham, Acernese,
  Ackley, Adams, Adhikari, Adya, Affeldt, Agathos, {et~al.}}]{GW1}
Abbott, R., Abbott, T., Abraham, S., {et~al.} 2020{\natexlab{a}}, \prl, 125,
  101102

\bibitem[{Abbott {et~al.}(2020{\natexlab{b}})Abbott, Abbott, Abraham, Acernese,
  Ackley, Adams, Adhikari, Adya, Affeldt, Agathos, {et~al.}}]{GW3}
Abbott, R., Abbott, T., Abraham, S., {et~al.} 2020{\natexlab{b}}, \prd, 102,
  043015

\bibitem[{Abuter {et~al.}(2020)Abuter, Amorim, Baub{\"o}ck, Berger, Bonnet,
  Brandner, Cardoso, Cl{\'e}net, De~Zeeuw, Dexter, {et~al.}}]{genzel3}
Abuter, R., Amorim, A., Baub{\"o}ck, M., {et~al.} 2020, \aap, 636, L5

\bibitem[{Ball {et~al.}(2005)Ball, Greenhill, Moran, Zaw, \& Henkel}]{ball}
Ball, G.~H., Greenhill, L.~J., Moran, J.~M., Zaw, I., \& Henkel, C. 2005, in
  Future Directions in High Resolution Astronomy, eds. J. Romney, \& M. Reid,
  ASP Conf. Ser., 340, 235

\bibitem[{Braatz {et~al.}(2010)Braatz, Reid, Humphreys, Henkel, Condon, \&
  Lo}]{MCP2}
Braatz, J., Reid, M., Humphreys, E., {et~al.} 2010, \apj, 718, 657

\bibitem[{Claussen {et~al.}(1984)Claussen, Heiligman, \&
  Lo}]{claussen1984water}
Claussen, M., Heiligman, G., \& Lo, K. 1984, \nat, 310, 298

\bibitem[{Darling(2017)}]{Darling_2017}
Darling, J. 2017, The Astrophysical Journal, 837, 100

\bibitem[{Davis \& Scrimgeour(2014)}]{Composicion}
Davis, T.~M. \& Scrimgeour, M.~I. 2014, \mnras, 442, 1117

\bibitem[{Do {et~al.}(2019)Do, Hees, Ghez, Martinez, Chu, Jia, Sakai, Lu,
  Gautam, O’neil, {et~al.}}]{Ghez3}
Do, T., Hees, A., Ghez, A., {et~al.} 2019, Science, 365, 664

\bibitem[{Dressler {et~al.}(1989)Dressler, Osterbrock, \&
  Miller}]{dressler1989active}
Dressler, A., Osterbrock, D., \& Miller, J. 1989, in IAU Symp, Vol. 134, 217

\bibitem[{EHT~Collaboration {et~al.}(2019)EHT~Collaboration, Akiyama, Alberdi,
  Alef, Asada, AZULY, {et~al.}}]{M87}
EHT~Collaboration, E. H.~T., Akiyama, K., Alberdi, A., {et~al.} 2019, ApJL,
  875, L1

\bibitem[{Einstein(1915)}]{einstein1915}
Einstein, A. 1915, Deutsche Akademie der Wissenschaften zu Berlin, Berlin, 844

\bibitem[{Ferrarese \& Merritt(2000)}]{ferrarese2000fundamental}
Ferrarese, L. \& Merritt, D. 2000, \apj, 539, L9

\bibitem[{Gebhardt {et~al.}(2000)Gebhardt, Bender, Bower, Dressler, Faber,
  Filippenko, Green, Grillmair, Ho, Kormendy,
  {et~al.}}]{gebhardt2000relationship}
Gebhardt, K., Bender, R., Bower, G., {et~al.} 2000, \apj, 539, L13

\bibitem[{Genzel \& Downes(1977)}]{genzel1977h2o}
Genzel, R. \& Downes, D. 1977, A\&AS, 30, 145

\bibitem[{Genzel {et~al.}(1997)Genzel, Eckart, Ott, \& Eisenhauer}]{genzel}
Genzel, R., Eckart, A., Ott, T., \& Eisenhauer, F. 1997, \mnras, 291, 219

\bibitem[{Genzel {et~al.}(2010)Genzel, Eisenhauer, \& Gillessen}]{genzel2}
Genzel, R., Eisenhauer, F., \& Gillessen, S. 2010, RMP, 82, 3121

\bibitem[{Ghez {et~al.}(1998)Ghez, Klein, Morris, \& Becklin}]{Ghez}
Ghez, A.~M., Klein, B., Morris, M., \& Becklin, E. 1998, ApJL, 509, 678

\bibitem[{Ghez {et~al.}(2008)Ghez, Salim, Weinberg, Lu, Do, Dunn, Matthews,
  Morris, Yelda, Becklin, {et~al.}}]{Ghez2}
Ghez, A.~M., Salim, S., Weinberg, N., {et~al.} 2008, \aj, 689, 1044

\bibitem[{Herrera-Aguilar \& Nucamendi(2015)}]{2015}
Herrera-Aguilar, A. \& Nucamendi, U. 2015, \prd, 92, 045024

\bibitem[{Herrnstein {et~al.}(2005)Herrnstein, Moran, Greenhill, \&
  Trotter}]{Hern}
Herrnstein, J., Moran, J.~M., Greenhill, L.~J., \& Trotter, A.~S. 2005, \apj,
  629, 719

\bibitem[{Kerr(1963)}]{Kerr}
Kerr, R.~P. 1963, \prl, 11, 237

\bibitem[{Koekemoer {et~al.}(1995)Koekemoer, Henkel, Greenhill, Dey, van
  Breugel, Codella, \& Antonucci}]{FirstTXS}
Koekemoer, A.~M., Henkel, C., Greenhill, L.~J., {et~al.} 1995, \nat, 378, 697

\bibitem[{Kormendy \& Ho(2013)}]{M-L}
Kormendy, J. \& Ho, L.~C. 2013, \araa, 51, 511

\bibitem[{Kormendy \& Richstone(1995)}]{kormendy1995inward}
Kormendy, J. \& Richstone, D. 1995, \araa, 33, 581

\bibitem[{Kuo {et~al.}(2011)Kuo, Braatz, Condon, Impellizzeri, Lo, Zaw,
  Schenker, Henkel, Reid, \& Greene}]{MCP3}
Kuo, C., Braatz, J., Condon, J., {et~al.} 2011, \apj, 727, 20

\bibitem[{Kuo {et~al.}(2018)Kuo, Constantin, Braatz, Chung, Witherspoon, Pesce,
  Impellizzeri, Gao, Hao, Woo, {et~al.}}]{kuo2018}
Kuo, C., Constantin, A., Braatz, J., {et~al.} 2018, \apj, 860, 169

\bibitem[{Marconi \& Hunt(2003)}]{marconi2003relation}
Marconi, A. \& Hunt, L.~K. 2003, \apj, 589, L21

\bibitem[{Nucamendi {et~al.}(2021)Nucamendi, Herrera-Aguilar, Lizardo-Castro,
  \& L{\'{o}}pez-Cruz}]{Herrera_ngc}
Nucamendi, U., Herrera-Aguilar, A., Lizardo-Castro, R., \& L{\'{o}}pez-Cruz, O.
  2021, ApJL, 917, L14

\bibitem[{Reid {et~al.}(2009)Reid, Braatz, Condon, Greenhill, Henkel, \&
  Lo}]{reid2009megamaser}
Reid, M., Braatz, J., Condon, J., {et~al.} 2009, \apj, 695, 287

\bibitem[{Rindler(1982)}]{RindlerSR1989}
Rindler, W. 1982, Introduction to special relativity (Oxford University Press)

\bibitem[{Schwarzschild(1916)}]{Schwa}
Schwarzschild, K. 1916, Sitzungsberichte der K{\"o}niglich Preu{\ss}ischen
  Akademie der Wissenschaften (Berlin, 189

\bibitem[{Surcis {et~al.}(2020)Surcis, Tarchi, \& Castangia}]{Gigamaser}
Surcis, G., Tarchi, A., \& Castangia, P. 2020, \aap, 637, A57

\bibitem[{Taylor {et~al.}(2002)Taylor, Peck, Henkel, Falcke, Mundell, arteand
  S.~A.~Baum, \& Gallimore}]{Taylor_2002}
Taylor, G.~B., Peck, A.~B., Henkel, C., {et~al.} 2002, \apj, 574, 88

\end{thebibliography}
\bibliographystyle{aa}

\end{document}